\documentclass[aps,showpacs]{revtex4}
\usepackage{graphicx}
\usepackage{dcolumn}
\usepackage{bm}
\newcommand{\beq}{\begin{equation}}
\newcommand{\eeq}{\end{equation}}
\newcommand{\bea}{\begin{eqnarray}}
\newcommand{\eea}{\end{eqnarray}}
\newcommand{\bec}{\begin{center}}
\newcommand{\enc}{\end{center}}
\newcommand{\alp}{\alpha}
\newcommand{\om}{\omega}
\newcommand{\Om}{\Omega}

\newcommand{\G}{\Gamma}
\newcommand{\s}{\sigma}

\newcommand{\tih}{\widetilde{h}}
\newcommand{\tiv}{\widetilde{v}}

\newcommand{\la}{\langle}
\newcommand{\ra}{\rangle}

\begin{document}
\title{
Deterministic photon-photon $\sqrt{\rm SWAP}$ gate 
using a lambda system
}
\author{Kazuki Koshino$^{1,2}$, Satoshi Ishizaka$^{3,4}$ 
and Yasunobu Nakamura$^{3,4,5}$}
\affiliation{
$^1$~College of Liberal Arts and Sciences, Tokyo Medical and Dental University,
2-8-30 Konodai, Ichikawa 272-0827, Japan
\\
$^2$~PRESTO, Japan Science and Technology Agency, 
4-1-8 Honcho, Kawaguchi 332-0012, Japan
\\
$^3$~Nano Electronics Research Laboratories,
NEC Corporation,
34 Miyukigaoka, Tsukuba 305-8501, Japan
\\
$^4$~INQIE, the University of Tokyo, 4-6-1 Komaba,
Meguro-ku, Tokyo 153-8505, Japan
\\
$^5$~The Institute of Physical and Chemical Research (RIKEN), 2-1 Hirosawa, Wako 351-0198, Japan 
}
\date{\today}
\begin{abstract}
We theoretically present a method to realize 
a deterministic photon-photon $\sqrt{\rm SWAP}$ gate
using a three-level lambda system 
interacting with single photons in reflection geometry. 
The lambda system is used completely passively 
as a temporary memory for a photonic qubit; 
the initial state of the lambda system may be arbitrary,
and active control by auxiliary fields is unnecessary
throughout the gate operations.
These distinct merits make this entangling gate 
suitable for deterministic and scalable quantum computation.
\end{abstract}
\maketitle

Single photons are promising candidates for implementing qubits in quantum computation
due to their long coherence times. 
Furthermore, one-qubit gates such as Hadamard and NOT gates
can be readily realized using linear optical elements~\cite{GK}.
Photonic qubits have the disadvantage that it
is difficult to realize two-qubit controlled gates 
such as controlled-NOT gates
due to the weak mutual interaction between photons~\cite{Mil}.
This problem has been partially overcome by 
linear optics quantum computation,
which enables probabilistic controlled gates 
that successfully operate depending on 
the measurement results of ancillary photons~\cite{KLM,OB}.

In the quest for realizing deterministic controlled gates in quantum optics,
a measurable nonlinear phase shift between single photons has been demonstrated
using a cavity quantum electrodynamics system
in the bad-cavity regime~\cite{Tur}.
This system has the characteristic that 
radiation from the atom is nearly completely forwarded
to a one-dimensional field
that is determined by the radiation pattern of the cavity.
Such one-dimensional configurations 
can be realized by a variety of physical systems,
including a leaky resonator interacting with an atom or a quantum dot~\cite{PB2,dot1},
a single emitter near a surface plasmon~\cite{pla},
and a superconducting qubit near a transmission line or a resonator~\cite{cc1,Fan}.
Since the incident light inevitably interferes with 
the radiation from the system due to the reduced dimensionality,
the effective light-matter interaction can be 
drastically enhanced under this configuration. 
Utilizing this property,
several quantum devices have been proposed to date,
such as controlled logic gates~\cite{Duan1,Duan2},
quantum-state converters~\cite{Cirac,Sham,Ima} 
and entanglers of photonic or material qubits~\cite{Ima,Hu1,Hu2}.
These devices perform their tasks
with the help of active quantum control of the material part
(such as initialization~\cite{Duan1,Duan2,Cirac,Sham,Ima,Hu1,Hu2}, 
single-qubit rotation~\cite{Duan1,Duan2}, 
and classical pumping~\cite{Cirac,Sham,Ima}) 
and by measurements~\cite{Hu1,Hu2}.

In the present study,
we theoretically point out a unique potential of a three-level lambda system
coupled to a one-dimensional photon field in the reflection geometry.
(A lambda system is hereafter referred to as an ^^ ^^ atom'',
although it can be implemented by other physical systems
such as semiconductor quantum dots 
and superconducting Josephson junctions~\cite{lam1,lam2,lam3}.)
We show that a deterministic photon-photon $\sqrt{\rm SWAP}$ gate 
can be realized by using the atom completely {\it passively} 
as a temporary memory for photonic qubits;
the initial state of the atom may be arbitrary including even mixed states,
and active control of the atom is unnecessary throughout successive gate operations.
These properties are quite advantageous 
when constructing scalable quantum networks.
Furthermore, since the $\sqrt{\rm SWAP}$ gate 
constitutes a universal set of quantum gates 
together with one-qubit gates~\cite{rtSWAP1,rtSWAP2},
the proposed scheme provides a vivid blueprint 
for future quantum computation,
that is deterministic and scalable.

\begin{figure}
\includegraphics[scale=1.2]{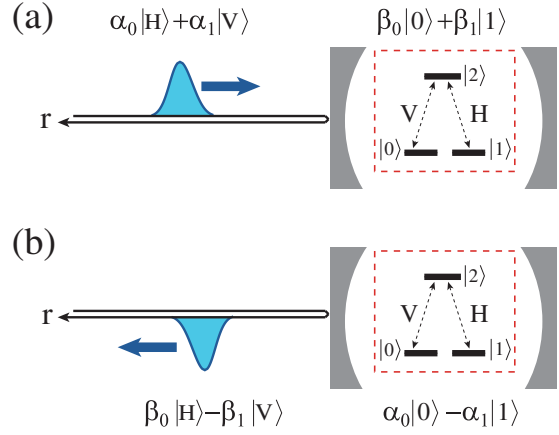}
\caption{\label{fig:1}
Interaction between a lambda system 
and a single photon propagating in one dimension. 
(a)~Initial state. 
The photonic and atomic qubits may be in arbitrary states.
(b)~Final state.
The lambda system is completely de-excited through radiative decay.
The photonic and atomic qubits can be completely swapped 
under appropriate conditions.
}\end{figure}

The physical setup considered in this study
is schematically illustrated in Fig.~\ref{fig:1}.
The atom has two degenerate ground states ($|0\ra$ and $|1\ra$) 
and an excited state ($|2\ra$),
and the transition frequency is $\Om$.
The $|1\ra\leftrightarrow|2\ra$ and $|0\ra\leftrightarrow|2\ra$
transitions in the atom are
assisted respectively by horizontally (H) and vertically (V)
polarized photons 
and the radiative decay rates for the $|2\ra\to|1\ra$ 
and $|2\ra\to|0\ra$ transitions are $\G_H$ and $\G_V$.
The total Hamiltonian including the atom and the photon field is given,
under the rotating-wave approximation, by
(putting $\hbar=c=1$)
\bea
{\cal H} &=& \Om \s_{22}
+\int dk \left[
k h_k^{\dagger} h_k
+i\sqrt{\frac{\G_H}{2\pi}}(\s_{21}h_k-h_k^{\dagger}\s_{12})
\right] \nonumber
\\
& &
+\int dk \left[
k v_k^{\dagger}v_k
+i\sqrt{\frac{\G_V}{2\pi}}(\s_{20}v_k-v_k^{\dagger}\s_{02})
\right],
\label{eq:H}
\eea
where 
$\s_{ij}(=|i\ra\la j|)$ is the atomic transition operator,
and $h_k$ ($v_k$) is the annihilation operator 
for the H (V) polarized photon with wave number $k$. 
As shown in Fig.~\ref{fig:1},
we define the spatial coordinate $r$  
along the propagation direction of the photon,
and assign the negative (positive) region to the input (output) ports. 
The real-space representation of the field operator $\tih_r$
is defined as the Fourier transform of $h_k$ 
by $\tih_r=(2\pi)^{-1/2}\int dk e^{ikr}h_k$.

The initial states of the photon and the atom are given by
$\alp_0|H\ra+\alp_1|V\ra$ and $\beta_0|0\ra+\beta_1|1\ra$, respectively
[Fig.~\ref{fig:1}(a)].
We denote the wave packet of the input photon in the real-space representation by $f(r)$,
which is normalized as $\int dr |f(r)|^2=1$.
The four basis states of the input are then given, 
in the multimode notation, by
\bea
|H,0\ra &=& \int dr f(r)\tih_r^{\dagger}|0\ra,
\label{eq:H0in}
\\
|H,1\ra &=& \int dr f(r)\tih_r^{\dagger}|1\ra,
\\
|V,0\ra &=& \int dr f(r)\tiv_r^{\dagger}|0\ra,
\\
|V,1\ra &=& \int dr f(r)\tiv_r^{\dagger}|1\ra.
\label{eq:V1in}
\eea
The output states are determined by the Schr\"odinger equations,
$|H,0\ra \to e^{-i{\cal H}t}|H,0\ra$ etc,
where ${\cal H}$ is the Hamiltonian of Eq.~(\ref{eq:H})
and the final time $t$ is a sufficiently large time
at which the atom is completely de-excited [Fig.~\ref{fig:1}(b)].
The time evolutions of $|H,0\ra$ and $|V,1\ra$ are trivial,
since the input photon does not interact with the atom 
and therefore propagates freely.
In contrast, the time evolutions of $|H,1\ra$ and $|V,0\ra$ are nontrivial, 
since the input photon may 
interact with the atom in these cases.
The output state vectors are given by
\bea
|H,0\ra &\to& \int dr g_1(r,t)\tih_r^{\dagger}|0\ra,
\label{eq:H0out1}
\\
|H,1\ra &\to& 
\int dr g_3(r,t)\tih_r^{\dagger}|1\ra
-\int dr g_2(r,t)\tiv_r^{\dagger}|0\ra,
\\
|V,0\ra &\to& 
\int dr g_4(r,t)\tiv_r^{\dagger}|0\ra
-\int dr g_2(r,t)\tih_r^{\dagger}|1\ra,
\\
|V,1\ra &\to& \int dr g_1(r,t)\tiv_r^{\dagger}|1\ra,
\label{eq:V1out1}
\eea
where $g_1$, $g_2$, $g_3$ and $g_4$ are determined by
\bea
g_1(r,t) &=& f(r-t),
\\
g_2(r,t) &=& \sqrt{\G_H\G_V}s(t-r),
\label{eq:g2}
\\
g_3(r,t) &=& f(r-t)-\G_H s(t-r),
\label{eq:g3}
\\
g_4(r,t) &=& f(r-t)-\G_V s(t-r),
\label{eq:g4}
\eea
where $s(t)$ is the atomic coherence induced by the input photon,
which evolves as
\beq
\frac{d}{dt}s(t)=\left(-i\Om-\frac{\G_H+\G_V}{2}\right)s(t)+f(-t).
\label{eq:s}
\eeq
These equations are derived in Appendix~\ref{sec:appa}
The probabilities of the occurrence and absence of 
the $|H,1\ra \leftrightarrow |V,0\ra$ transition 
are quantified by
$P=\int dr |g_2(r,t)|^2$ and $P'=\int dr |g_3(r,t)|^2$=$\int dr |g_4(r,t)|^2$,
which satisfy the sum rule of $P+P'=1$.

We here consider a case in which the pulse length $l$ 
of the input photon is sufficiently long to satisfy $l \gg \G_{H,V}^{-1}$.
In this case, Eq.~(\ref{eq:s}) can be solved adiabatically.
Denoting the detuning of the input photon by $\om$
[namely, $f(r)\sim e^{i(\Om+\om)r}$],
$s(t)$ is given by
$s(t)=\frac{2}{\G_H+\G_V-2i\om}f(-t)$.
Substituting this into Eqs.~(\ref{eq:H0out1})--(\ref{eq:g4})
and neglecting the translational motion of the photon,
the four basis states 
are transformed as follows on reflection:
\bea
|H,0\ra &\to& |H,0\ra,
\label{eq:H0out2}
\\
|H,1\ra &\to& 
{\textstyle
\frac{\G_V-\G_H-2i\om}{\G_H+\G_V-2i\om}|H,1\ra
-\frac{2\sqrt{\G_H\G_V}}{\G_H+\G_V-2i\om}|V,0\ra,
}\label{eq:H1out2}
\\
|V,0\ra &\to& 
{\textstyle
-\frac{2\sqrt{\G_H\G_V}}{\G_H+\G_V-2i\om}|H,1\ra
+\frac{\G_H-\G_V-2i\om}{\G_H+\G_V-2i\om}|V,0\ra,
}\label{eq:V0out2}
\\
|V,1\ra &\to& |V,1\ra.
\label{eq:V1out2}
\eea
The case of $\G_H=\G_V$ is of particular interest as a quantum logic gate.
When the input photon is in resonance with the atom ($\om=0$),
this gate behaves as an atom-photon SWAP gate.
As illustrated in Fig.~\ref{fig:1},
the quantum states of atomic and photonic qubits are exchanged on reflection as
\beq
(\alp_0|H\ra+\alp_1|V\ra)(\beta_0|0\ra+\beta_1|1\ra)
\to
(\beta_0|H\ra-\beta_1|V\ra)(\alp_0|0\ra-\alp_1|1\ra).
\label{eq:swap}
\eeq
Thus, the atom functions as a quantum memory of the photonic qubit,
which is indispensable for long-distance quantum key distribution
using quantum repeaters.
On the other hand, when the detuning of the input photon is set 
to the linewidth of the atom ($\om=\pm\G_H$),
this gate behaves as an atom-photon $\sqrt{\rm SWAP}$ gate.
For example, when $\om=-\G_H$, 
$|H,1\ra \to 2^{-1/2}(e^{i\pi/4}|H,1\ra+e^{3i\pi/4}|V,0\ra)$ and 
$|V,0\ra \to 2^{-1/2}(e^{3i\pi/4}|H,1\ra+e^{i\pi/4}|V,0\ra)$,
whereas $|H,0\ra$ and $|V,1\ra$ remain unchanged.

\begin{figure}
\includegraphics[scale=0.5]{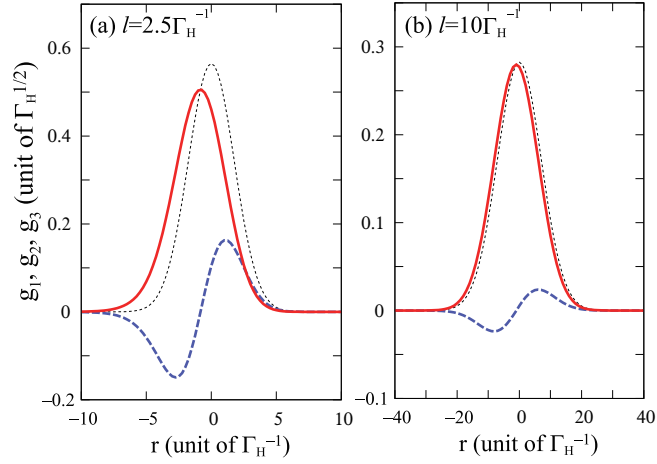}
\caption{\label{fig:2}
Shapes of the output wave packets,
$g_1$ (thin dotted line), $g_2$ (solid line) and $g_3$ (dashed line),
for the case of the atom-photon SWAP gate ($\G_H=\G_V$ and $\om=0$).
The natural phase factor $e^{i\Om(r-t)}$ is removed. 
The pulse length is (a)~$l=2.5\G_H^{-1}$ and (b)~$l=10\G_H^{-1}$. 
}\end{figure}

\begin{figure}
\includegraphics[scale=0.5]{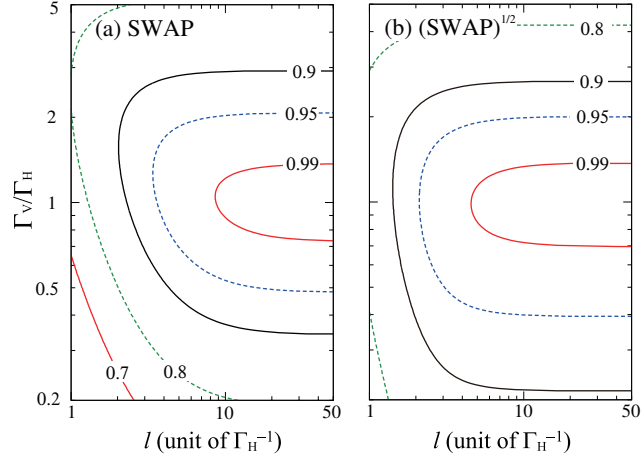}
\caption{\label{fig:3}
Contour plots of the average gate fidelities 
for the atom-photon (a)~SWAP ($\om=0$) and 
(b)~$\sqrt{\rm SWAP}$ ($\om=-\G_H$) gates,
as functions of the pulse length $l$ and $\G_V/\G_H$.
}\end{figure}

To observe the effects of a finite pulse length $l$,
the shapes of $g_1$, $g_2$ and $g_3$(=$g_4$) are plotted in Fig.~\ref{fig:2}
for the case of the atom-photon SWAP gate ($\G_H=\G_V$ and $\om=0$).
The input mode function is assumed to be Gaussian,
$f(r)=(2/\pi l^2)^{1/4}\exp(-r^2/l^2+i\Om r)$. 
It is observed that
$g_2$ is slightly delayed relative to $g_1$
due to absorption and re-emission by the atom. 
The delay time is of the order of $\G_H^{-1}$.
However, this delay becomes negligible when the input pulse is long 
($l\gg\G_H^{-1}$) as in Fig.~\ref{fig:2}(b). 
$g_2$ becomes almost identical to $g_1$ whereas $g_3$ vanishes.
The average gate fidelities of the atom-photon SWAP and $\sqrt{\rm SWAP}$ gates 
are respectively given by~\cite{avgf1,avgf2}
\bea
\bar{F}_{\rm SWAP} &=& \frac{1+\left|1+ \int dr g_2^{\ast}g_1 \right|^2}{5},
\\
\bar{F}_{\sqrt{\rm SWAP}} &=& \frac{1+\left| 1+\frac{1+i}{4} 
\int dr (g_3^{\ast}+g_4^{\ast}-2ig_2^{\ast})g_1 \right|^2}{5}.
\eea
In Fig.~\ref{fig:3},
$\bar{F}_{\rm SWAP}$ and $\bar{F}_{\sqrt{\rm SWAP}}$
are plotted as functions of the pulse length $l$ 
and the ratio $\G_V/\G_H$ of the atomic decay rates.
The conditions for achieving high-fidelity operations are given 
by $l \gg \G_{H,V}^{-1}$ and $\G_H/\G_V \simeq 1$ for both gates.

\begin{figure}
\includegraphics[scale=1.3]{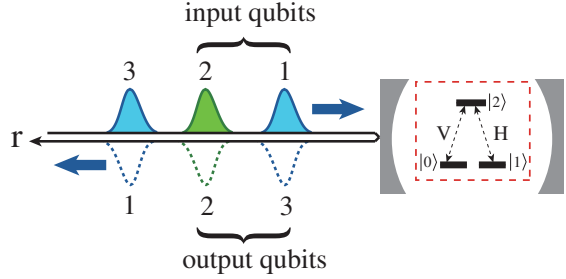}
\caption{\label{fig:4}
Illustration of the photon-photon $\sqrt{\rm SWAP}$ gate.
Photons 1 and 3 are in resonance with the atom ($\om=0$),
whereas photon 2 is detuned ($\om=\pm\G_H$).
The input qubits are the polarization states of photons 1 and 2,
whereas the output qubits are those of photons 3 and 2.
}\end{figure}

These atom--photon gates 
are highly useful for a variety of purposes in quantum information processing.
As an illuminative example, we show that 
a photon-photon $\sqrt{\rm SWAP}$ gate
that operates deterministically and with a high gate fidelity
can be realized by using an atom as a temporary quantum memory.
This fact implies that 
deterministic all-optical quantum computation is possible,
since a $\sqrt{\rm SWAP}$ gate constitutes a universal set of quantum gates
together with one-photon gates,
which can be realized by linear optical elements.
Figure~\ref{fig:4} shows a schematic illustration of the photon-photon $\sqrt{\rm SWAP}$ gate.
Three photons (1, 2, and 3) are forwarded to the atom with sufficiently large time intervals between them. 
The initial state of the atom may be an arbitrary superposition 
of the two ground states, $|0\ra$ and $|1\ra$.
Photons 1 and 3 are in resonance with the atom ($\om=0$), 
whereas photon 2 is slightly detuned ($\om=\pm\G_H$). 
We assign photons 1 and 2 as the input qubits,
and photons 3 and 2 as the output qubits.
We can then confirm the following $\sqrt{\rm SWAP}$ operation
(for example, for $\om=-\G_H$ for photon 2):
\bea
|H\ra_1|H\ra_2 &\to& |H\ra_3|H\ra_2,
\label{eq:HH}
\\
|H\ra_1|V\ra_2 &\to& 
2^{-1/2}(e^{i\pi/4}|H\ra_3|V\ra_2+e^{-i\pi/4}|V\ra_3|H\ra_2)
\label{eq:HV}
\\
|V\ra_1|H\ra_2 &\to& 
2^{-1/2}(e^{-i\pi/4}|H\ra_3|V\ra_2+e^{i\pi/4}|V\ra_3|H\ra_2)
\label{eq:VH}
\\
|V\ra_1|V\ra_2 &\to& |V\ra_3|V\ra_2.
\label{eq:VV}
\eea
The initial states of the atom and photon 3,
both of which may be arbitrary, are respectively transferred
to the final states of photon 1 and the atom in the following manner:
\bea
\alp_0|0\ra_a + \alp_1|1\ra_a & \to & \alp_0|H\ra_1 - \alp_1|V\ra_1,
\\
\beta_0|H\ra_3 + \beta_1|V\ra_3 & \to & \beta_0|0\ra_a - \beta_1|1\ra_a,
\eea
where the subscript ^^ ^^ $a$'' denotes the atom.
These states are unentangled with the output qubits
and therefore do not affect the gate.
They can also be recycled for subsequent gate operations.
Four comments on this gate are in order.
(i)~This gate enables the $\sqrt{\rm SWAP}$ operation between two photons
having different frequencies.
The SWAP operation between them can be achieved by using this gate twice.
Therefore, the current scheme can be extended to construct
the $\sqrt{\rm SWAP}$ operation between two photons
having {\it the same} frequency.
(ii)~The present scheme does not depend on 
optical nonlinearity nor interference between single photons.
Therefore, the proposed gate operates with a high fidelity
irrespective of the pulse shapes and time intervals of the input photons, 
provided the pulses are sufficiently long.
This implies that high stability of optical paths,
which is essential in many optical experiments,
is not required.
(iii)~The atom may be in an arbitrary de-excited state initially.
Even if the atom is in a mixed state,
it can be restored to a pure state automatically by the first input photon.
Therefore, the quantum coherence of the atom should be maintained
only during the three photons interact with the atom.
(iv)~Throughout successive gate operations, 
there is no need for auxiliary control fields 
to manipulate the atomic quantum state.
Namely, the atom is used completely passively
as a temporary quantum memory.
These merits make the proposed scheme 
quite advantageous for constructing a scalable quantum network.

Finally, we estimate the effects of practical noises and imperfections 
such as radiative loss, finite spin-coherence times,
discrepancy between $\G_H$ and $\G_V$,
and finite pulse lengths,
assuming that the lambda system is implemented 
by a charged quantum dot in a photonic crystal nanocavity.
The typical values of the cavity-QED parameters 
are $(g, \gamma, \kappa)\sim 2\pi\times(16, 0.2, 32)$~GHz,
and therefore $\Gamma(\sim g^2/\kappa)\sim 2\pi\times 8$~GHz.
Then, the photon loss rate is estimated at 
$\gamma/(\Gamma+\gamma)\sim$2.5\% per one gate operation.
The gate fidelity can be estimated with a help of Fig.~3;
when the pulse length $l$ is 400~ps (20$\Gamma^{-1}$)
and $\G_H/\G_V=1.4$ for example,
the fidelity of the photon-photon $\sqrt{\rm SWAP}$ gate 
becomes $(0.99)^3\sim 0.97$.
The time intervals between photons should be shorter 
than the homogeneous spin-coherence time 
of the order of $\mu$s~\cite{spin1,spin2}.

In summary, we have investigated the interaction 
between a three-level lambda system and a single photon
propagating in one dimension (Fig.~\ref{fig:1}),
and observed that this atom--photon system behaves 
as SWAP and $\sqrt{\rm SWAP}$ gates
when the two decay rates in the atom are close ($\G_H\simeq\G_V$).
Furthermore, successive input of three photons enables
a photon-photon $\sqrt{\rm SWAP}$ gate (Fig.~\ref{fig:4})
that can operate deterministically. 
The distinct advantage of the proposed gate
is that the atomic qubit is used completely passively; 
the atomic qubit may be in an arbitrary initial state,
and any active control of the atomic qubit is unnecessary
throughout the gate operations.
Therefore, the proposed gate is suitable 
for constructing scalable quantum networks and computers.

The authors are grateful to T. Yamamoto, T. Kato, R. Shimizu, 
and N. Matsuda for fruitful discussions.
This research was partially supported by 
the Nakajima Foundation,
MEXT KAKENHI (17GS1204, 21104507),
the Special Coordination Funds for Promoting Science and Technology, 
and the CREST program of the Japan Science and Technology Agency (JST).

\appendix
\section{Derivation of Eqs.~(10)--(14)}
\label{sec:appa}
\subsection{Heisenberg equations}
From the Hamiltonian of Eq.~(1),
the Heisenberg equations for $h_k$ and $\s_{12}$ are given by
\bea
\frac{d}{d\tau}h_k &=& -ikh_k-\sqrt{\frac{\G_H}{2\pi}}\s_{12},
\label{eq:hk}
\\
\frac{d}{d\tau}\s_{12} &=& -i\Om\s_{12}
+\sqrt{\G_H}(\s_{11}-\s_{22})\tih_0+\sqrt{\G_V}\s_{10}\tiv_0,
\label{eq:s12'}
\eea
where $\tih_0$ is the real-space representation 
of the field operator at $r=0$, namely,
$\tih_0=(2\pi)^{-1/2}\int dk h_k$.
The initial and final moments are respectively set at $\tau=0$ and $t$.
Then, from Eq.~(\ref{eq:hk}), 
$h_k(\tau)$ ($0<\tau<t$) is represented in two ways,
\bea
h_k(\tau) &=& h_k(0)e^{-ik\tau}-\sqrt{\frac{\G_H}{2\pi}}
\int_0^{\tau}d\tau'\s_{12}(\tau')e^{-ik(\tau-\tau')},
\\
h_k(\tau) &=& h_k(t)e^{-ik(\tau-t)}+\sqrt{\frac{\G_H}{2\pi}}
\int_{\tau}^t d\tau'\s_{12}(\tau')e^{-ik(\tau-\tau')}.
\eea
As the Fourier transform of the above equations,
$\tih_0(\tau)$ is given by
\bea
\tih_0(\tau) &=& \tih_{-\tau}(0)-\frac{\sqrt{\G_H}}{2}\s_{12}(\tau),
\label{eq:tih0}
\\
\tih_0(\tau) &=& \tih_{t-\tau}(t)+\frac{\sqrt{\G_H}}{2}\s_{12}(\tau).
\eea
Equating the right-hand sides, 
introducing a new label $r(=t-\tau)$,
and using the symmetry of the system,
we obtain the following set of equations:
\bea
\tih_r(t) &=& \tih_{r-t}(0)-\sqrt{\Gamma_H}\s_{12}(t-r),
\label{eq:hrt}
\\
\tiv_r(t) &=& \tiv_{r-t}(0)-\sqrt{\Gamma_V}\s_{02}(t-r),
\label{eq:vrt}
\eea
where $0<r<t$. These equations are known as the input-output relation.
Substituting Eq.~(\ref{eq:tih0}) 
into Eq.~(\ref{eq:s12'}) and using the symmetry, we obtain
\bea
\frac{d}{d\tau}\s_{12} &=& 
\left(-i\Om-\frac{\G_H+\G_V}{2}\right)\s_{12}
+\sqrt{\G_H}(\s_{11}-\s_{22})\tih_{-\tau}(0)
+\sqrt{\G_V}\s_{10}\tiv_{-\tau}(0),
\label{eq:s12}
\\
\frac{d}{d\tau}\s_{02} &=& 
\left(-i\Om-\frac{\G_H+\G_V}{2}\right)\s_{02}
+\sqrt{\G_V}(\s_{00}-\s_{22})\tiv_{-\tau}(0)
+\sqrt{\G_H}\s_{01}\tih_{-\tau}(0).
\label{eq:s02}
\eea

\subsection{Temporal evolution of $|V,0\ra$}
We investigate the temporal evolution of the input states, Eqs.~(2)--(5).
As a nontrivial case, we consider here the evolution of $|V,0\ra$.
From Eqs.~(4) and (8), 
the input and output state vectors are written as
\bea
|\varphi_{in}\ra &=& \int dr f(r) \tiv_r^{\dagger}|0\ra,
\label{eq:vpi}
\\
|\varphi_{out}\ra &=& \int dr g_4(r;t) \tiv_r^{\dagger}|0\ra
- \int dr g_2(r;t) \tih_r^{\dagger}|1\ra.
\label{eq:vpo}
\eea
These two state vectors are related by 
$|\varphi_{out}\ra=e^{-i{\cal H}t}|\varphi_{in}\ra$.
The following two properties are useful in the arguments below:
(I)~$e^{-i{\cal H}t}|0\ra=|0\ra$ and $e^{-i{\cal H}t}|1\ra=|1\ra$
since ${\cal H}|0\ra={\cal H}|1\ra=0$.
(II)~$\tih_r(0)|\varphi_{in}\ra=0$ and 
$\tiv_r(0)|\varphi_{in}\ra=f(r)|0\ra$,
since the field commutators are given by
$[\tiv_r, \tiv_{r'}^{\dagger}]=\delta(r-r')$
and $[\tih_r, \tiv_{r'}^{\dagger}]=0$.

From Eq.~(\ref{eq:vpo}) and the property~(I), 
$g_2$ and $g_4$ are determined by
\bea
g_2(r,t) &=& -\la 1|\tih_r|\varphi_{out}\ra
=-\la 1|\tih_r(t)|\varphi_{in}\ra,
\\
g_4(r,t) &=& \la 0|\tiv_r|\varphi_{out}\ra
=\la 0|\tiv_r(t)|\varphi_{in}\ra,
\eea
where $A(t)=e^{i{\cal H}t}Ae^{-i{\cal H}t}$ (Heisenberg representation).
Using Eqs.~(\ref{eq:hrt}), (\ref{eq:vrt}), (\ref{eq:vpi}) and the property~(II), 
$g_2$ and $g_4$ are recast into the following forms:
\bea
g_2(r,t) &=& 
\sqrt{\G_H}\la 1|\s_{12}(t-r)|\varphi_{in}\ra,
\\
g_4(r,t) &=& f(r-t)-\sqrt{\G_V}\la 0|\s_{02}(t-r)|\varphi_{in}\ra.
\eea
$s_{02}(\tau)\equiv\la 0|\s_{02}(\tau)|\varphi_{in}\ra$ and 
$s_{12}(\tau)\equiv\la 1|\s_{12}(\tau)|\varphi_{in}\ra$
represent the atomic coherence induced by the input photon.
Their equations of motion are given,
from Eqs.~(\ref{eq:s12})--(\ref{eq:vpi}) and the property~(II), by
\bea
\frac{d}{d\tau}s_{02}(\tau) &=& 
\left(-i\Om-\frac{\G_H+\G_V}{2}\right)s_{02}(\tau)
+\sqrt{\G_V}\la 0|\s_{00}(\tau)-\s_{22}(\tau)|0\ra f(-\tau),
\\
\frac{d}{d\tau}s_{12}(\tau) &=& 
\left(-i\Om-\frac{\G_H+\G_V}{2}\right)s_{12}(\tau)
+\sqrt{\G_V}\la 1|\s_{10}(\tau)|0\ra f(-\tau).
\eea
Since $\la 0|\s_{00}(\tau)-\s_{22}(\tau)|0\ra=\la 1|\s_{10}(\tau)|0\ra=1$
from the property~(I),
$s_{02}(\tau)$ and $s_{12}(\tau)$ becomes identical. 
Introducing $s=s_{02}/\sqrt{\G_V}=s_{12}/\sqrt{\G_V}$, 
we obtain 
\bea
g_2(r,t) &=& 
\sqrt{\G_H\G_V}s(t-r),
\\
g_4(r,t) &=& f(r-t)-\G_V s(t-r),
\\
\frac{d}{d\tau}s(\tau) &=& 
\left(-i\Om-\frac{\G_H+\G_V}{2}\right)s(\tau)+f(-\tau).
\eea
Thus, Eqs.~(11), (13) and (14) of the main text are derived.


\end{document}